\documentclass[pra,superscriptaddress,showpacs,twocolumn,10pt]{revtex4}

\usepackage{amssymb}
\usepackage{amsmath}
\usepackage{graphicx}

\newtheorem{lemma}{Lemma}
\newtheorem{theorem}{Theorem}

\newtheorem{example}{Example}

\begin{document}

\title{Perfect many-to-one teleportation with stabilizer states}
\author{Guoming Wang}
\email{wgm00@mails.tsinghua.edu.cn}
\author{Mingsheng Ying}
\email{yingmsh@tsinghua.edu.cn}

\affiliation{State Key Laboratory of Intelligent Technology and
Systems, Department of Computer Science and Technology, Tsinghua
University, Beijing, China, 100084}

\date{\today}

\begin{abstract}
We study the possibility of performing
perfect teleportation of unknown quantum states
from multiple senders to a single receiver with a previously
shared stabilizer state.
In the model we considered, the utilized stabilizer state is
partitioned into several subsystems and
then each subsystem is distributed to a distinct party.
We present two sufficient conditions for a stabilizer state to
achieve a given nonzero teleportation capacity with respect to a given
partition plan. The corresponding teleportation protocols are also
explicitly given. Interestingly, we find that even mixed stabilizer states are
also useful for perfect many-to-one teleportation. Finally,
our work provides a new perspective from stabilizer formalism to
view the standard teleportation protocol and also suggests a new technique
for analyzing teleportation capability of multipartite entangled states.
\end{abstract}


\maketitle

\section{Introduction}

Entanglement is an intrigue feature of quantum mechanics. It has
been exploited as a resource to carry out various amazing tasks
which are impossible in classical physics. A remarkable example is
quantum teleportation \cite{BB93} which allows indirect
transmission of quantum information between distant parties by
using previously shared entanglement and classical communication
between them. Indeed, teleportation has become a basic building
block of many quantum communication and quantum computation
protocols nowadays.

It is widely acknowledged that a thorough understanding of the
power of entanglement in information procession is one of the
major goals of quantum information theory. One of the key steps
toward this goal is to give a complete characterization of
teleportation capability of quantum entanglement. However, up to
now, most progress in this direction is restricted to the simple
case of bipartite entangled states
\cite{BB93,HH99,LK00,B00,RH01,SL01,AF02,GK02,VV03,Y03,RD03,KC04,G04,JF05,
LC05,GR06,BS06,M06}. Results about the power of general
multipartite entangled states in teleportation are still scarce.
This is partially due to the exponentially growing complexity of
multipartite states. To avoid such an unmanageable complexity,
some authors chose to consider special multipartite states that
own certain symmetry. For examples, one can see Refs.
\cite{LM02,FL03,LJ05,R05,YC06,CZ06,AP06,MX07,LJ07}.

In this paper we study the usefulness of stabilizer states for
perfect teleportation. Stabilizer states have played an important
role in quantum information theory, especially in the field of
quantum error correction \cite{S95,S96} and cluster state quantum
computation \cite{RB01}. They can be described in an elegant and
compact form named the stabilizer formalism \cite{G96,G97}, which
has also lead to novel perspectives to many phenomena in quantum
information science and quantum mechanics \cite{TG05,ND07,WY07}.

Let us first fix the model of teleportation in the multipartite
case. Suppose $\rho$ is an $n$-qudit state. Divide its $n$ qudits
into $m$ groups $T_1,T_2,\dots,T_m$ for some $2 \le m \le n$ and
distribute the subsystem $T_i$ to the $i$-th party $A_i$, for
$i=1,2,\dots,m$. Now assume that $A_i$ has an unknown $a_i$-qudit
state $\sigma_i$, for $i=1,2,\dots,m-1$. We want to know whether
$A_1,A_2,\dots,A_{m-1}$ can \textit{simultaneously faithfully}
teleport the states $\sigma_1,\sigma_2,\dots,\sigma_{m-1}$ to
$A_m$ by performing local operations on the particles they have
and classical communications among them(LOCC). If this is
possible, then $(a_1,a_2,\dots,a_{m-1})$ is called an achievable
teleportation capacity for $\rho$ with respect to the grouping
plan $T_1,T_2,\dots,T_{m}$. For a given state $\rho$, each
grouping strategy will define a region of achievable teleportation
capacities. Our question is exactly to determine such a region for
all possible grouping strategies.

Our main results are two sufficient conditions for a stabilizer
state to achieve a given nonzero teleportation capacity with
respect to a given partition plan. While the first condition is
only suitable for bipartitions, the second can be applied to
general partitions. The corresponding teleportation protocols are
also explicitly given. Interestingly,
we find that even mixed stabilizer states are also useful
for perfect many-to-one teleportation. Finally, our work
provides a new perspective from the stabilizer formalism to
view the standard teleportation protocol and also
suggests a new technique for analyzing teleportation capability of
multipartite entangled states.

This paper is organized as follows. In Sec. II we briefly recall
some basic facts about the stabilizer formalism. In Sec. III, we
study the usefulness of stabilizer states for perfect
teleportation and also construct our teleportation protocols. In
Sec. IV, we analyze several concrete examples by applying our
theorems. Finally, Sec. V summarizes our results.

\section{Preliminary}

In this section, we review some fundamental facts about stabilizer
state and its corresponding stabilizer formalism. Although in most
literatures the notion of stabilizer state was put forward in the
context of multiqubit systems, it can actually be generalized
without essential difficulty to arbitrary higher-dimensional
systems as well. Similar topics have also been explored in Refs.
\cite{G98,NB02,Y02,HD05}. So here we directly start with the
general higher-dimensional case.

Consider a $d$-dimensional Hilbert space. Define
$X_{(d)}$ and $Z_{(d)}$ as follows:
\begin{equation}
\begin{array}{l}
X_{(d)}|j\rangle=|j \oplus 1\rangle,\\
Z_{(d)}|j\rangle={\omega^{j}|j\rangle},
\end{array}
\label{equ:pauli}
\end{equation}
where $j \in \mathbb{Z}_d$, $\omega=e^{i\frac{2\pi}{d}}$ is the $d$-th root of unity
over the complex field and the `$\oplus$' sign denotes addition
modulo $d$. Then the matrices $\{X_{(d)}^aZ_{(d)}^b:a,b=
0,1,\dots,d-1\}$ are considered as the generalized Pauli matrices
over the $d$-dimensional space. In what follows, without causing
ambiguity, we will omit the subscript `$(d)$' and use $X,Z$ to
denote $X_{(d)},Z_{(d)}$. The commutation relations among the
generalized Pauli matrices over the $d$-dimensional space are
given by
\begin{equation}
(X^aZ^b)(X^jZ^k)=\omega^{bj-ak}(X^jZ^k)(X^aZ^b).
\label{equ:commuterelation}
\end{equation}
It can be checked that if $d$ is even and $ab$ is odd, the
eigenvalues of $X^aZ^b$ are $\omega^{1/2}, \omega^{c+1/2},
\omega^{2c+1/2}, \dots, \omega^{d-c+1/2}$ for some factor $c$ of
$d$; otherwise, the eigenvalues of $X^aZ^b$ are $1, \omega^c,
\omega^{2c}, \dots, \omega^{d-c}$ for some factor $c$ of $d$.

Let $X_i,Z_i$ denote the operations of $X,Z$ on the $i$-th qudit
respectively. The generalized Pauli group on $n$ qudits
$G^{(d)}_n$ is generated under multiplication by the Pauli
matrices acting on each qudit, together with the phase factor
$\gamma=\sqrt{\omega}$, i.e.
\begin{equation}
\begin{array}{l}
G^{(d)}_n=\langle \gamma, X_1, Z_1, X_2, Z_2, \dots, X_n,
Z_n\rangle. \label{equ:pauligroup}
\end{array}
\end{equation}
By Eq.(\ref{equ:commuterelation}), for any
$g=\gamma^c X^{a_1}Z^{b_1}\otimes X^{a_2}Z^{b_2} \otimes
\dots \otimes X^{a_n}Z^{b_n}$, $h=\gamma^{c'}
X^{a'_1}Z^{b'_1}\otimes X^{a'_2}Z^{b'_2}\otimes\dots \otimes
X^{a'_n}Z^{b'_n} \in G^{(d)}_n$, their commutation relation is
given by
\begin{equation}
gh=\omega^{\sum\limits_{i=1}^{n}(b_ia'_i-a_ib'_i)}hg.
\end{equation}
In particular, $g$ and $h$ commute if and only if
${\sum_{i=1}^{n}(b_ia'_i-a_ib'_i)}$ is a multiple of $d$.

For a set of commuting operators $g_1,g_2,\dots,g_k \in G^{(d)}_n$,
we say that they are independent if $\forall i=1,2,\dots,k$,
\begin{equation}
\langle g_1,g_2,\dots,g_k\rangle \neq \langle g_1,g_2,\dots,g_{i-1},g_{i+1},\dots,
g_k\rangle.
\end{equation}
Define $G'^{(d)}_n$ to be the subset of $G^{(d)}_n$ composed of
all the operators whose eigenvalues are of the form $1, \omega^c,
\omega^{2c}, \dots, \omega^{d-c}$ for some factor $c$ of $d$. Now
suppose $g_1,g_2,\dots,g_k$ are independent commuting operators in $G'^{(d)}_n$. Let
\begin{equation}
S=\langle g_1,g_2,\dots,g_k \rangle
\end{equation}
be the Abelian subgroup generated by them. A state $|\psi\rangle$
is said to be stabilized by $S$, or $S$ is the stabilizer of
$|\psi\rangle$, if
\begin{equation}\begin{array}{ll}
g_i|\psi\rangle=|\psi\rangle, & \forall i=1,2,\dots,k.
\end{array}\end{equation}
All the states stabilized by $S$ constitute a subspace denoted by
$V_S$. With the fact
$\sum_{j=0}^{d-1}{\omega^{j\lambda}}=0,\forall
\lambda=1,2,\dots,d-1$, one can verify that the projection
operator onto $V_S$ is
\begin{equation}
P_{S}=\frac{1}{d^k}\prod\limits_{i=1}^{k}(\sum\limits_{j=0}^{d-1}{g^j_i}).
\label{equ:ps}
\end{equation}
Then the maximally mixed state over $V_S$ is
\begin{equation}
\rho_S=P_S/tr(P_S). \label{equ:rhos}
\end{equation}
In particular, if there is a unique pure state (up to an overall phase)
stabilized by $S$, then $g_1,g_2,\dots,g_k$ are called a
\textit{complete} set of stabilizer generators and $S$ is called a
\textit{complete} stabilizer.

In practice we are often interested in the stabilized subspace
$V_S$, which is the simultaneous eigenspace of the operators
$g_1,g_2,\dots,g_k$ corresponding to the eigenvalues
$1,1,\dots,1$. But in general we can also consider the
simultaneous eigenspace of $g_1,g_2,\dots,g_k$ corresponding to
their other eigenvalues. In what follows, we will use
$P{(g_1, g_2, \dots, g_k; \overrightarrow{x})}$ to denote the
projection operator onto the simultaneous eigenspace of
$g_1,g_2,\dots,g_k$ corresponding to the eigenvalues
$\omega^{x_1},\omega^{x_2},\dots,\omega^{x_k}$, where
$\overrightarrow{x}=(x_1, x_2, \dots,x_k) \in \mathbb{Z}^{k}_d$.
With the fact $\sum_{j=0}^{d-1}{\omega^{j\lambda}}=0,\forall
\lambda=1,2,\dots,d-1$, one can see
\begin{equation}
P{(g_1, g_2, \dots, g_k; \overrightarrow{x})}=
\frac{1}{d^k}\prod\limits_{i=1}^{k}({{\sum\limits_{j=0}^{d-1}{\omega^{-jx_i}g^{j}_i}}}),
\label{equ:projection}
\end{equation}
In particular, $P_S=P{(g_1, g_2, \dots, g_k;
\overrightarrow{0})}$.

For any two subgroups $H_1,H_2$ of $G^{(d)}_{n}$,
if there exists a bijective map $N:H_1
\rightarrow H_2$ such that for any $h_1,h_2 \in H_1$,
$N(h_1h_2)=N(h_1)N(h_2)$, then
we say $H_1$ and $H_2$ are isomorphic. We will denote
this isomorphism by $H_1 \cong H_2$.
Given several operators $g_1,g_2,\dots,g_k \in
G^{(d)}_n$, we are usually interested in the commutation relations
among them. In this situation, we may write, e.g.
$g_1=\overline{Z}_1$, $g_2=\overline{X}_1$, $g_3=\overline{Z}_2$,
$g_4=\overline{Z}_3$. The intention of this writing is to indicate
that $\langle g_1,g_2,g_3,g_4 \rangle \cong \langle
Z_1,X_1, Z_2,Z_3 \rangle$ and the isomorphism between them is
induced by $N(g_1)=Z_1$, $N(g_2)=X_1$, $N(g_3)=Z_2$, $N(g_4)=Z_3$.
Note that $g_1$ may not actually be the action of $Z$ on the first
qudit, and similarly for $g_2$, $g_3$, $g_4$.

\section{Perfect teleportation with stabilizer states}

In this section we study the usefulness of the state $\rho_S$
given by Eqs.(\ref{equ:ps}) and (\ref{equ:rhos}) for perfect
teleportation with multiple senders and one receiver. Note that
only when $S$ is a complete stabilizer, $\rho_S$ is a pure state.
In other cases, $\rho_S$ is a mixed state. But our discussion
below does not need to discriminate between the two cases because
it essentially does not depend on the purity of $\rho_S$.

At first, we need to introduce several definitions and notations.
We will use $[1,n]$ to denote the set of integers
$\{1,2,\dots,n\}$. If $T_1,T_2,\dots,T_l$ are disjoint proper
subsets of $[1,n]$ and they satisfy $\cup_{i=1}^{l} T_i=[1,n]$,
then we say $\{T_1,T_2,\dots,T_l\}$ is a partition of $[1,n]$. For any $T \subset [1,n]$,
we use $|T|$ to denote the number of elements in $T$ and
also use $T^C$ to denote the complement of $T$ in $[1,n]$.
For any $T \subset [1,n]$ and $g=\gamma^c X^{a_1}Z^{b_1} \otimes
X^{a_2}Z^{b_2}\otimes \dots \otimes X^{a_n}Z^{b_n} \in G^{(d)}_n$,
define the restriction of $g$ on $T$ to be
\begin{equation}
g^{(T)}=\bigotimes\limits_{i \in T} X^{a_i}Z^{b_i}.
\end{equation}
Furthermore, for $S=\langle g_1,g_2,\dots,g_k\rangle$,
define the restriction of $S$ on $T$ to be
\begin{equation}
S^{(T)}=\langle \gamma,g^{(T)}_1,g^{(T)}_2,\dots,g^{(T)}_k\rangle.
\end{equation}
One can easily see that the choice of
stabilizer generators $g_1,g_2,\dots,g_k$ does not affect the result
$S^{(T)}$. So it is well-defined.
In addition, in what follows, we will
use the subset $T=\{i_1,i_2,\dots,i_t\} \subset [1,n]$
to represent the subsystem of $\rho_S$ composed of
the $i_1$-th, $i_2$-th, ..., $i_t$-th qudits.
We also use $\rho^{(T)}_S$ to denote
the reduced density matrix of $\rho_S$ on this subsystem.
Finally, for several subgroups
$P,P_1,P_2,\dots,P_k$ of $G^{(d)}_n$, if we write
\begin{equation}
P=\prod_{i=1}^{k}{P_i}=P_1P_2\dots P_k
\end{equation}
we mean that each element of $P_i$ commutes with
each element of $P_j$, $\forall 1 \le i \neq j \le k$,
and
\begin{equation}
P=\{g_1g_2\dots g_k: \forall i=1,2,\dots,k, g_i \in P_i\}.
\end{equation}

Now let us reformulate our problem precisely. Suppose
$g_1,g_2,\dots,g_k$ are independent commuting operators in
$G'^{(d)}_n$. The state $\rho_S$ given by Eqs.(\ref{equ:ps}) and
(\ref{equ:rhos}) is the maximally mixed state over the subspace
stabilized by $S=\langle g_1,g_2,\dots,g_k \rangle$. Assume that
$\{T_1,T_2,\dots,T_{m+1}\}$ is a partition of $[1,n]$.
$A_1,A_2,\dots,A_{m+1}$ are distant parties and $A_i$ holds the
subsystem $T_i$ of $\rho_S$, for $i=1,2,\dots,m+1$. Now suppose
$A_i$ has an unknown $a_i$-qudit state $\sigma_i$, for
$i=1,2,\dots,m$. If $A_1,A_2,\dots,A_m$ can \textit{simultaneously
faithfully} teleport the states
$\sigma_1,\sigma_2,\dots,\sigma_{m}$ to $A_{m+1}$ by performing
LOCC operations on the particles they have, the
$(a_1,a_2,\dots,a_m)$ is said to be an achievable teleportation
capacity for $\rho_S$ with respect to $\{T_1,T_2,\dots,T_{m+1}\}$.
Our goal is to determine the region of achievable teleportation capacities
for $\rho_S$ with respect to an arbitrary partition plan.

Before presenting our main theorems, it is necessary to prove a
lemma at first.

In Ref.\cite{BD06} the authors found an interesting theorem which
states that for any two isomorphic subgroups $G$ and $H$ of the
Pauli group on $n$ qubits, there exists a unitary operation $U$
such that for any $g \in G$, there exists $h \in H$ such that
$g=UhU^{\dagger}$ up to an overall phase. Here our lemma can
be viewed as a partial extension of this theorem to the higher
dimensional case.

\begin{lemma}
If a subgroup $H$ of $G^{(d)}_{n}$ is isomorphic to $G=\langle
\gamma^{c}, Z^{a_1}_1,Z^{a_2}_2,\dots,Z^{a_s}_s,X^{b_1}_1,X^{b_2}_2,
\dots,X^{b_t}_t\rangle$ for some $t \le s \le n$,
$c \in \mathbb{Z}_{2d}$ and $a_1, a_2, \dots,
a_s, b_1, b_2, \dots, b_t \in \mathbb{Z}_d$, then there exists a
unitary operation $U$ such that for any $h \in H$, there exists
$g \in G$ such that $h=UgU^{\dagger}$.
\end{lemma}

\textit{Proof:} By the definition of isomorphism we can write $H$
as
\begin{equation}
H=\langle \gamma^c, \overline{Z}^{a_1}_1, \overline{Z}^{a_2}_2, \dots,
\overline{Z}^{a_s}_s, \overline{X}^{b_1}_1, \overline{X}^{b_2}_2,
\dots, \overline{X}^{b_t}_t \rangle
\label{equ:HH}
\end{equation}
for some $\overline{Z}_1,
\overline{Z}_2, \dots, \overline{Z}_s, \overline{X}_1,
\overline{X}_2, \dots, \overline{X}_t \in G^{(d)}_{n}$.

Note that $\overline{Z}_1, \overline{Z}_2, \dots, \overline{Z}_s$
mutually commute and their simultaneous eigenspace corresponding
to the eigenvalues $\omega^{x_1},\omega^{x_2},\dots,\omega^{x_s}$
is $d^{n-s}$-dimensional, $\forall x_1, x_2,
\dots, x_s \in \mathbb{Z}_d$. For any
$\overrightarrow{x}=(x_1,x_2,\dots,x_s) \in \mathbb{Z}^s_d$,
define $\overrightarrow{x}|_{[t+1,s]}=(x_{t+1},x_{t+2},\dots,x_s)$.
Suppose
$\{|\overline{\psi}(\overrightarrow{x}|_{[t+1,s]};\alpha)\rangle\}_{\alpha=1}^{d^{n-s}}$
is an arbitrary orthonormal basis of the simultaneous eigenspace
of $\overline{Z}_1, \overline{Z}_2, \dots,
\overline{Z}_t, \overline{Z}_{t+1}, \dots, \overline{Z}_s$
corresponding to the eigenvalues $1, 1, \dots, 1,
\omega^{x_{t+1}},\omega^{x_{t+2}},\dots,\omega^{x_s}$. Define
\begin{equation}
|\overline{\phi}(\overrightarrow{x};\alpha)\rangle=
\overline{X}^{x_1}_1\overline{X}^{x_2}_2\dots\overline{X}^{x_t}_t
|\overline{\psi}(\overrightarrow{x}|_{[t+1,s]};\alpha)\rangle,
\end{equation}
$\forall \alpha=1,2,\dots,d^{n-s}$. Then
$\{|\overline{\phi}(\overrightarrow{x};\alpha)\}_{\alpha=1}^{d^{n-s}}$
is an orthonormal basis for the simultaneous
eigenspace of $\overline{Z}_1, \overline{Z}_2, \dots,
\overline{Z}_s$ corresponding to the eigenvalues
$\omega^{x_1},\omega^{x_2},\dots,\omega^{x_s}$. To see this, one
only needs to realize that for $\forall i \in [1,t]$,
\begin{equation}\begin{array}{l}
\overline{Z}_i|\overline{\phi}(\overrightarrow{x};\alpha)\rangle\\
=\overline{Z}_i\overline{X}^{x_1}_1\overline{X}^{x_2}_2\dots\overline{X}^{x_t}_t
|\overline{\psi}(\overrightarrow{x}|_{[t+1,s]};\alpha)\rangle\\
=\omega^{x_i}\overline{X}^{x_1}_1\overline{X}^{x_2}_2\dots\overline{X}^{x_t}_t\overline{Z}_i
|\overline{\psi}(\overrightarrow{x}|_{[t+1,s]};\alpha)\rangle\\
=\omega^{x_i}\overline{X}^{x_1}_1\overline{X}^{x_2}_2\dots\overline{X}^{x_t}_t
|\overline{\psi}(\overrightarrow{x}|_{[t+1,s]};\alpha)\rangle\\
=\omega^{x_i}|\overline{\phi}(\overrightarrow{x};\alpha)\rangle,
\end{array}\end{equation}
and $\forall i \in [t+1,s]$,
\begin{equation}\begin{array}{l}
\overline{Z}_i|\overline{\phi}(\overrightarrow{x};\alpha)\rangle\\
=\overline{Z}_i\overline{X}^{x_1}_1\overline{X}^{x_2}_2\dots\overline{X}^{x_t}_t
|\overline{\psi}(\overrightarrow{x}|_{[t+1,s]};\alpha)\rangle\\
=\overline{X}^{x_1}_1\overline{X}^{x_2}_2\dots\overline{X}^{x_t}_t\overline{Z}_i
|\overline{\psi}(\overrightarrow{x}|_{[t+1,s]};\alpha)\rangle\\
=\omega^{x_i}\overline{X}^{x_1}_1\overline{X}^{x_2}_2\dots\overline{X}^{x_t}_t
|\overline{\psi}(\overrightarrow{x}|_{[t+1,s]};\alpha)\rangle\\
=\omega^{x_i}|\overline{\phi}(\overrightarrow{x};\alpha)\rangle.
\end{array}\end{equation}

Similarly, suppose
$\{|\psi(\overrightarrow{x}|_{[t+1,s]};\alpha)\rangle\}_{\alpha=1}^{d^{n-s}}$
is an arbitrary orthonormal basis of
the simultaneous eigenspace
of ${Z}_1, {Z}_2, \dots,
{Z}_t, {Z}_{t+1}, \dots, {Z}_s$
corresponding to the eigenvalues $1, 1, \dots, 1,
\omega^{x_{t+1}},\omega^{x_{t+2}},\dots,\omega^{x_s}$.
Define
\begin{equation}
|{\phi}(\overrightarrow{x};\alpha)\rangle=
{X}^{x_1}_1{X}^{x_2}_2\dots{X}^{x_t}_t
|{\psi}(\overrightarrow{x}|_{[t+1,s]};\alpha)\rangle,
\end{equation}
$\forall \alpha=1,2,\dots,d^{n-s}$. Then
$\{|\phi(\overrightarrow{x};\alpha)\}_{\alpha=1}^{d^{n-s}}$
is an orthonormal basis for the simultaneous
eigenspace of ${Z}_1, {Z}_2, \dots,
{Z}_s$ corresponding to the eigenvalues
$\omega^{x_1},\omega^{x_2},\dots,\omega^{x_s}$,
$\forall \overrightarrow{x}=(x_1,x_2,\dots,x_s) \in \mathbb{Z}^s_d$.

Define the following unitary operation
\begin{equation}
U=\sum\limits_{\overrightarrow{x} \in \mathbb{Z}^s_d}\sum\limits_{\alpha=1}^{d^{n-s}}
{|\overline{\phi}(\overrightarrow{x};\alpha)\rangle \langle
\phi(\overrightarrow{x};\alpha)|}.
\label{equ:unitary}
\end{equation}

From its definition, one can easily see that $U$ is indeed unitary
and
\begin{equation}\begin{array}{ll}
\overline{Z}_i=UZ_iU^{\dagger}, &\forall i=1,2,\dots,s.
\label{equ:zz}
\end{array}\end{equation}
Moreover, $\forall i=1,2,\dots,t$,
$\forall \overrightarrow{x}=(x_1,x_2,\dots,x_s) \in \mathbb{Z}^s_d$,
$\forall \alpha=1,2,\dots,d^{n-s}$, we have
\begin{equation}\begin{array}{l}
\overline{X}_i|\overline{\phi}(\overrightarrow{x};\alpha)\rangle\\
=\overline{X}_i\overline{X}^{x_1}_1\overline{X}^{x_2}_2\dots
\overline{X}^{x_t}_t|\overline{\psi}(\overrightarrow{x}|_{[t+1,s]};\alpha)\rangle\\
=\overline{X}^{x_1}_1\overline{X}^{x_2}_2\dots
\overline{X}^{x_i \oplus 1}_i\dots
\overline{X}^{x_t}_t|\overline{\psi}(\overrightarrow{x}|_{[t+1,s]};\alpha)\rangle\\
=|\overline{\phi}(\overrightarrow{x} \oplus \overrightarrow{e}_i;\alpha)\rangle,
\end{array}\end{equation}
and
\begin{equation}\begin{array}{l}
UX_iU^{\dagger}|\overline{\phi}(\overrightarrow{x};\alpha)\rangle\\
=UX_i|\phi(\overrightarrow{x};\alpha)\rangle\\
=UX_iX^{x_1}_1X^{x_2}_2\dots X^{x_t}_t|\psi(\overrightarrow{x}|_{[t+1,s]};\alpha)\rangle\\
=UX^{x_1}_1X^{x_2}_2\dots X^{x_i \oplus 1}_i \dots X^{x_t}_t|\psi(\overrightarrow{x}|_{[t+1,s]};\alpha)\rangle\\
=U|\phi(\overrightarrow{x} \oplus \overrightarrow{e}_i;\alpha)\rangle\\
=|\overline{\phi}(\overrightarrow{x} \oplus \overrightarrow{e}_i;\alpha)\rangle,
\end{array}\end{equation}
where $\overrightarrow{e}_i=(0,\dots,0,1,0,\dots,0)$ ($1$
is the $i$-th element) and `$\oplus$' denotes addition modulo $d$.
Since $\{\overline{\phi}(\overrightarrow{x};\alpha)\}_{\overrightarrow{x} \in \mathbb{Z}^s_d,1 \le \alpha \le d^{n-s}}$
is an orthonormal basis of the $n$-qudit Hilbert space,
two above equations actually tell us that
\begin{equation}\begin{array}{ll}
\overline{X}_i=UX_iU^{\dagger}, &\forall i=1,2,\dots,t.
\label{equ:xx}
\end{array}\end{equation}

Now the validity of this lemma follows immediately
from Eqs.(\ref{equ:zz}) and (\ref{equ:xx}).

\hfill $\blacksquare$

With the help of this lemma, we find that for a
bipartition $\{T_1,T_2\}$, the structure of
$S^{(T_2)}$ can influence the
teleportation capacity of $\rho_S$ with respect to this partition
$\{T_1,T_2\}$, as the following theorem states:

\begin{theorem}
Suppose $\{T_1,T_2\}$ is a bipartition of $[1,n]$. If there exist
subgroups $P_1$ and $P_2$ of $S$ such that
\begin{equation}\begin{array}{l}
S^{(T_2)}=P^{(T_2)}_1P^{(T_2)}_2,\\
P^{(T_2)}_1\cong G^{(d)}_t,\\
P^{(T_2)}_2\cong\langle \gamma, Z^{a_{1}}_{1},Z^{a_{2}}_{2},
\dots,Z^{a_{s}}_s, X^{b_{1}}_{1},X^{b_2}_{2},\dots,X^{b_{u}}_{u}\rangle,
\label{equ:t1}
\end{array}\end{equation}
for some $t \ge 0$, $s \ge u \ge 0$, and
$a_1,a_2,\dots,a_s, b_1,b_2$, $\dots,b_u \in \mathbb{Z}_d$,
then $t$ is an achievable teleportation capacity for $\rho_S$
with respect to the partition $\{T_1,T_2\}$.
\end{theorem}

\textit{Proof:} Suppose $|T_1|=m$ and $|T_2|=n-m$. By
Eq.(\ref{equ:t1}) we can find independent generators
$g_1,g_2,\dots,g_k$ of $S$ such that
\begin{equation}\begin{array}{ll}
g_{2i-1}=R_{2i-1} \otimes \overline{Z}_i,& \forall 1 \le i \le t;\\
g_{2i}=R_{2i} \otimes \overline{X}_i,& \forall 1 \le i \le t;\\
g_{2t+i}=R_{2t+i} \otimes \overline{Z}^{a_i}_{t+i},& \forall 1 \le i \le s;\\
g_{2t+s+i}=R_{2t+s+i} \otimes \overline{X}^{b_i}_{t+i},& \forall 1 \le i \le u;\\
g_{i}=R_{i} \otimes I, & \forall 2t+s+u+1 \le i \le k.\\
\label{equ:g}
\end{array}\end{equation}
where $R_1,R_2,\dots,R_k$ are some operators on the subsystem $T_1$,
$\overline{Z}_1,\overline{Z}_2,\dots,\overline{Z}_{t+s},
\overline{X}_1,\overline{X}_2,\dots,\overline{X}_{t+u} \in G^{(d)}_{n-m}$
are operators on the subsystem $T_2$.

By lemma 1, we can find a unitary operator $U$ acting on the subsystem $T_2$
such that
\begin{equation}\begin{array}{ll}
U\overline{Z}_iU^{\dagger}=Z_i,& \forall 1 \le i \le t+s,\\
U\overline{X}_iU^{\dagger}=X_i,& \forall 1 \le i \le t+u.\\
\label{equ:gt}
\end{array}\end{equation}

Define
\begin{equation}\begin{array}{ll}
h_i=(I \otimes U)g_i(I \otimes U^{\dagger}),
\end{array}\end{equation}
$\forall i=1,2,\dots,k$.
Then we have
\begin{equation}\begin{array}{ll}
h_{2i-1}=R_{2i-1} \otimes {Z}_i,& \forall 1 \le i \le t;\\
h_{2i}=R_{2i} \otimes {X}_i,& \forall 1 \le i \le t;\\
h_{2t+i}=R_{2t+i} \otimes {Z}^{a_i}_{t+i},& \forall 1 \le i \le s;\\
h_{2t+s+i}=R_{2t+s+i} \otimes {X}^{b_i}_{t+i},& \forall 1 \le i \le u;\\
h_{i}=R_{i} \otimes I, & \forall 2t+s+u+1 \le i \le k.\\
\label{equ:h}
\end{array}\end{equation}

Suppose
\begin{equation}\begin{array}{l}
T_2=\{i_1,i_2,\dots,i_{n-m}\}
\end{array}\end{equation}
with $i_1<i_2<\dots<i_{n-m}$. One can see
Eq.(\ref{equ:h}) implies $t \le n-m$. So
define
\begin{equation}\begin{array}{l}
T'_2=\{i_1,i_2,\dots,i_t\},\\
T''_2=\{i_{t+1},i_{t+2},\dots,i_{n-m}\}.\\
\end{array}\end{equation}

Since $g_1,g_2,\dots,g_{k}$ mutually commute,
by the definition of $h_1,h_2,\dots,h_{k}$
we know that they also mutually commute.
For $i=1,2,\dots,t$, define
\begin{equation}\begin{array}{ll}
h'_{2i-1}&=R_{2i-1} \otimes h^{(T'_2)}_{2i-1}\\
&=R_{2i-1} \otimes Z_{i},\\
h'_{2i}&=R_{2i} \otimes h^{(T'_2)}_{2i}\\
&=R_{2i} \otimes X_{i},\\
\label{equ:h'}
\end{array}\end{equation}
Then $h'_1, h'_2, \dots, h'_{2t}$ are commuting operators
on the subsystem $T_1 \cup T'_2$.

Now suppose Alice and Bob hold the
subsystems $T_1$ and $T_2$ of $\rho_S$ respectively.
Assume Alice has a $t$-qudit system $M$ and it is in an
unknown state $\sigma$. We firstly propose a teleportation
protocol, and then prove its validity. The protocol is as follows:

(1)Bob performs the unitary operation $U$ on his subsystem $T_2$.

(2)Alice performs the projective measurement consisting of the
projection operators
$\{P(h'_1,h'_2,\dots,h'_{2t};\overrightarrow{x}):\overrightarrow{x}
\in \mathbb{Z}^{2t}_d\}$ on her $T_1$ subsystem of $\rho_S$ and $M$.
Then she tells the measurement outcome
$\overrightarrow{x}=(x_1,x_2,\dots,x_{2t})$ to Bob.

(3)Bob performs the unitary operation
\begin{equation}
V(\overrightarrow{x})=\bigotimes\limits_{i=1}^{t}{(Z^{-x_{2i}}X^{x_{2i-1}})}
\label{equ:vx}
\end{equation}
on the subsystem $T'_2$.

Now we prove that after this procedure, the state of the subsystem
$T'_2$ is exactly $\sigma$.

After step (1), one can see that $\rho_S$ becomes
$\rho_{S'}=P_{S'}/tr(P_{S'})$ where $S'=\langle h_1, h_2, \dots,
h_k \rangle$ and $P_{S'}$ is the projection operator onto the
subspace stabilized by $S'$. By Eqs.(\ref{equ:ps}) and (\ref{equ:h}), we have
\begin{equation}\begin{array}{ll}
P_{S'}&=\frac{1}{d^k}\prod\limits_{i=1}^{k}(\sum\limits_{j=0}^{d-1}{h^j_i})\\
&=\frac{1}{d^k}\sum\limits_{j_1,\dots,j_k=0}^{d-1}\prod\limits_{i=1}^{k}{h^{j_i}_i}\\
&=\frac{1}{d^k}\sum\limits_{\overrightarrow{j} \in \mathbb{Z}^{k}_{d}}
A(\overrightarrow{j}) \otimes B(\overrightarrow{j})\otimes C(\overrightarrow{j})
\label{equ:ps'}
\end{array}\end{equation}
where $\overrightarrow{j}=(j_1,j_2,\dots,j_k)$, and
\begin{equation}\begin{array}{ll}
A(\overrightarrow{j})&=R^{j_1}_1R^{j_2}_2 \dots R^{j_k}_k,\\
B(\overrightarrow{j})&=\bigotimes\limits_{i=1}^{t}{(Z^{j_{2i-1}}X^{j_{2i}})},\\
C(\overrightarrow{j})&=\bigotimes\limits_{i=1}^{u}{(Z^{a_ij_{2t+i}}X^{b_ij_{2t+s+i}})}
\otimes \bigotimes\limits_{i=u+1}^{s}{Z^{a_ij_{2t+i}}} \otimes I\\
\label{equ:abc}
\end{array}\end{equation}
are operators on the subsystems $T_1$, $T'_2$ and $T''_2$ respectively.

Then, by Eqs.(\ref{equ:projection}) and (\ref{equ:h'}),
the projection operators of Alice's projective measurement in step (2) are
\begin{equation}\begin{array}{l}
P(h'_1,h'_2,\dots,h'_{2t};\overrightarrow{x})\\
=\frac{1}{d^{2t}}\prod\limits_{i=1}^{2t}({{\sum\limits_{j=0}^{d-1}{\omega^{-jx_i}h'^{j}_i}}})\\
=\frac{1}{d^{2t}}\sum\limits_{j_1,\dots,j_{2t}=0}^{d-1}\prod\limits_{i=1}^{2t}({{{\omega^{-j_ix_i}h'^{j_i}_i}}})\\
=\frac{1}{d^{2t}}
\sum\limits_{\overrightarrow{j} \in\mathbb{Z}^{2t}_{d}}
(\prod\limits_{i=1}^{2t}{\omega^{-j_ix_i}})
D(\overrightarrow{j}) \otimes E(\overrightarrow{j}),\\
\label{equ:pm}
\end{array}\end{equation}
where $\overrightarrow{j}=(j_1,j_2,\dots,j_{2t})$, and
\begin{equation}\begin{array}{ll}
D(\overrightarrow{j})&=R^{j_1}_1R^{j_2}_2 \dots R^{j_{2t}}_{2t},\\
E(\overrightarrow{j})&=\bigotimes\limits_{i=1}^{t}{(Z^{j_{2i-1}}X^{j_{2i}})}\\
\end{array}\end{equation}
are operators acting on the subsystems $T_1$ and $M$ respectively,
$\forall \overrightarrow{x}=(x_1,x_2,\dots,x_{2t}) \in \mathbb{Z}^{2t}_d$.

Since the density matrix of any $t$-qudit state can always be written
as the linear combination of the generalized Pauli group elements
$\{L(\overrightarrow{y})\equiv\bigotimes_{i=1}^t(Z^{y_{2i-1}}X^{y_{2i}}):
\overrightarrow{y}=(y_1,y_2,\dots,y_{2t}) \in
\mathbb{Z}^{2t}_{d}\}$, we can assume that the unknown state $\sigma$ is
\begin{equation}
\begin{array}{ll}
\sigma&=\sum\limits_{\overrightarrow{y} \in
\mathbb{Z}^{2t}_{d}}\lambda_{\overrightarrow{y}}L(\overrightarrow{y})\\
&=\sum\limits_{\overrightarrow{y} \in
\mathbb{Z}^{2t}_{d}}\lambda_{\overrightarrow{y}}
\bigotimes\limits_{i=1}^t(Z^{y_{2i-1}}X^{y_{2i}})
\label{equ:sigma}
\end{array}
\end{equation}
for some coefficients $\{\lambda_{\overrightarrow{y}}\}$.
Then after Alice's measurement in step (2), if the measurement outcome is
$\overrightarrow{x}=(x_1,x_2,\dots,x_{2t})$, the state of the
whole system becomes, up to a normalizing factor,
\begin{equation}\begin{array}{ll}
\widetilde{\rho}
&=P(h'_1,h'_2,\dots,h'_{2t};\overrightarrow{x})(P_{S'}\otimes \sigma)P(h'_1,h'_2,\dots,h'_{2t};\overrightarrow{x})\\
&=\frac{1}{d^{4t+k}}
\sum\limits_{\overrightarrow{j} \in \mathbb{Z}^{k}_{d}}
\sum\limits_{\overrightarrow{j}' \in \mathbb{Z}^{2t}_{d}}
\sum\limits_{\overrightarrow{j}'' \in \mathbb{Z}^{2t}_{d}}
\sum\limits_{\overrightarrow{y} \in \mathbb{Z}^{2t}_{d}}
[\lambda_{\overrightarrow{y}}
(\prod\limits_{i=1}^{2t}{\omega^{-x_i(j'_i+j''_i)}})\\
&F(\overrightarrow{j},\overrightarrow{j}',\overrightarrow{j}'')
\otimes
B(\overrightarrow{j})
\otimes
C(\overrightarrow{j})
\otimes
N(\overrightarrow{j}',\overrightarrow{j}'',\overrightarrow{y})]
\label{equ:intricate}
\end{array}\end{equation}
where $\overrightarrow{j}=(j_1,j_2,\dots,j_k)$,
$\overrightarrow{j}'=(j'_1,j'_2,\dots,j'_{2t})$,
$\overrightarrow{j}''=(j''_1,j''_2,\dots,j''_{2t})$, and
\begin{equation}
\begin{array}{l}
F(\overrightarrow{j},\overrightarrow{j}',\overrightarrow{j}'')\\
=D(\overrightarrow{j}')A(\overrightarrow{j})D(\overrightarrow{j}'')\\
=R^{j'_1}_1R^{j'_2}_2 \dots R^{j'_{2t}}_{2t}
R^{j_1}_1R^{j_2}_2 \dots R^{j_k}_k
R^{j''_1}_1R^{j''_2}_2 \dots R^{j''_{2t}}_{2t},\\
N(\overrightarrow{j}',\overrightarrow{j}'',\overrightarrow{y})\\
=E(\overrightarrow{j}')L(\overrightarrow{y})E(\overrightarrow{j}'')\\
=\bigotimes\limits_{i=1}^t (Z^{j'_{2i-1}}X^{j'_{2i}}Z^{y_{2i-1}}X^{y_{2i}}Z^{j''_{2i-1}}X^{j''_{2i}})\\
\end{array}
\label{equ:efgh}
\end{equation}
are operators on the subsystems $T_1$ and $M$
respectively.

Although Eqs.(\ref{equ:intricate}) and (\ref{equ:efgh}) seem
very intricate, after tracing out the subsystems $T_1$,
$T''_2$ and $M$, the reduced density matrix on the subsystem
$T'_2$ will become much simpler.
Let us consider each summation term
$F(\overrightarrow{j},\overrightarrow{j}',\overrightarrow{j}'')
\otimes B(\overrightarrow{j}) \otimes C(\overrightarrow{j})
\otimes N(\overrightarrow{j}',\overrightarrow{j}'',\overrightarrow{y})$.

Firstly, define
\begin{equation}\begin{array}{ll}
\Theta&=\{(j_{2t+1},j_{2t+2},\dots,j_{k}):\\
&\forall i=1,2,\dots,s, \hspace{8pt} a_ij_{2t+i}\equiv 0(mod \hspace{8pt} d);\\
&\forall i=1,2,\dots,u, \hspace{8pt} b_ij_{2t+s+i}\equiv 0(mod \hspace{8pt} d)\}.\\
\end{array}\end{equation}
Then $tr(C(\overrightarrow{j}))\neq 0$ if and only
if $(j_{2t+1},j_{2t+2},\dots,j_{k}) \in \Theta$.

Secondly,
$tr(N(\overrightarrow{j}',\overrightarrow{j}'',\overrightarrow{y}))\neq 0$
if and only if $\forall i=1,2,\dots,2t$, $y_{i}+j'_{i}+j''_{i}\equiv 0(mod$ $d)$;

Thirdly, note that
\begin{equation}\begin{array}{l}
F(\overrightarrow{j},\overrightarrow{j}',\overrightarrow{j}'')=
\omega^{\xi(\overrightarrow{j},\overrightarrow{j}',\overrightarrow{j}'')}
R^{j'_{1}+j_{1}+j''_{1}}_1
R^{j'_{2}+j_{2}+j''_{2}}_2
\dots\\
R^{j'_{2t}+j_{2t}+j''_{2t}}_{2t}
R^{j_{2t+1}}_{2t+1}R^{j_{2t+2}}_{2t+2} \dots R^{j_k}_k\\
\label{equ:fff}
\end{array}\end{equation}
for some $\xi(\overrightarrow{j},\overrightarrow{j}',\overrightarrow{j}'') \in \mathbb{Z}_d$.
Define
\begin{equation}\begin{array}{ll}
\Omega&=\{(j_{2t+1},j_{2t+2},\dots,j_{k}): \exists \lambda \neq 0\in \mathbb{C}, s.t.\\
&R^{j_{2t+1}}_{2t+1}R^{j_{2t+2}}_{2t+2}\dots R^{j_{k}}_{k}=\lambda I\}.
\end{array}\end{equation}
Now we will prove
$tr(F(\overrightarrow{j},\overrightarrow{j}',\overrightarrow{j}''))\neq 0$
if and only if $\forall i=1,2,\dots,2t$,
$j_{i}+j'_{i}+j''_{i}\equiv 0(mod$ $d)$ and
$(j_{2t+1},j_{2t+2},\dots,j_k) \in \Omega$.
To prove this, one needs to realize that
if $F(\overrightarrow{j},\overrightarrow{j}',\overrightarrow{j}'')=\mu I$
for some $\mu$, then it should commute with $R_i$, $\forall i=1,2,\dots,2t$.
Besides, since $g_1,g_2,\dots,g_k$ mutually commute, by Eqs.(\ref{equ:g}),
we get
\begin{equation}\begin{array}{ll}
R_{2i-1}R_{2i}=\omega^{-1} R_{2i}R_{2i-1}, &\forall 1 \le i \le t;\\
R_{2i-1}R_{2i'-1}=R_{2i'-1}R_{2i-1}, &\forall 1 \le i \neq i' \le t;\\
R_{2i-1}R_{2i'}=R_{2i'}R_{2i-1}, &\forall 1 \le i \neq i' \le t;\\
R_{2i}R_{2i'}=R_{2i'}R_{2i}, &\forall 1 \le i \neq i' \le t;\\
R_iR_j=R_jR_i, &\forall 1 \le i \le 2t, \forall 2t+1 \le j \le k.\\
\end{array}\end{equation}
So by Eq.(\ref{equ:fff})), $F(\overrightarrow{j},\overrightarrow{j}',\overrightarrow{j}'')$
commutes with $R_i$, $\forall i=1,2,\dots,2t$
if and only if
$j_{i}+j'_{i}+j''_{i}\equiv 0(mod$ $d)$, $\forall i=1,2,\dots,2t$.
In this case, Eq.(\ref{equ:fff}) reduces into
\begin{equation}\begin{array}{l}
F(\overrightarrow{j},\overrightarrow{j}',\overrightarrow{j}'')=
\omega^{\xi(\overrightarrow{j},\overrightarrow{j}',\overrightarrow{j}'')}
R^{j_{2t+1}}_{2t+1}R^{j_{2t+2}}_{2t+2} \dots R^{j_k}_k.\\
\end{array}\end{equation}
Then we have $tr(F(\overrightarrow{j},\overrightarrow{j}',\overrightarrow{j}'')) \neq 0$
if and only if $(j_{2t+1},j_{2t+2},\dots,j_k) \in \Omega$.

Summarizing the above argument, we know that only when $\forall i
=1,2,\dots,2t$, $y_i=j_i\equiv (-j'_i-j''_i)(mod$ $d)$ and
$(j_{2t+1},j_{2t+2},\dots,j_k) \in \Theta \cap \Omega$,
the corresponding term
$F(\overrightarrow{j},\overrightarrow{j}',\overrightarrow{j}'')
\otimes B(\overrightarrow{j}) \otimes C(\overrightarrow{j})
\otimes
N(\overrightarrow{j}',\overrightarrow{j}'',\overrightarrow{y})$
will not vanish after tracing out
$F(\overrightarrow{j},\overrightarrow{j}',\overrightarrow{j}'')$,
$C(\overrightarrow{j})$ and
$N(\overrightarrow{j}',\overrightarrow{j}'',\overrightarrow{y})$.
Note that by the definition of $\Theta$ and $\Omega$,
for any $(j_{2t+1},j_{2t+2},\dots,j_k) \in \Theta \cap \Omega$,
\begin{equation}\begin{array}{l}
\epsilon I=h^{j_{2t+1}}_{2t+1}h^{j_{2t+2}}_{2t+2}\dots h^{j_{k}}_{k}
\label{equ:zeta}
\end{array}\end{equation}
for some $\epsilon \in \mathbb{C}$. Suppose a state $|\psi\rangle$
is stabilized by $S'=\langle h_{1},h_{2},\dots,h_{k}\rangle$.
Then by Eq.(\ref{equ:zeta}) we obtain
\begin{equation}\begin{array}{l}
\epsilon |\psi\rangle=h^{j_{2t+1}}_{2t+1}h^{j_{2t+2}}_{2t+2}\dots h^{j_{k}}_{k}|\psi\rangle
=|\psi\rangle,\\
\end{array}\end{equation}
which is possible only if $\epsilon=1$. So for any
$(j_{2t+1},j_{2t+2},\dots,j_k) \in \Theta \cap \Omega$,
\begin{equation}\begin{array}{l}
I=h^{j_{2t+1}}_{2t+1}h^{j_{2t+2}}_{2t+2}\dots h^{j_{k}}_{k}
\end{array}\end{equation}
Therefore, when
$\forall i=1,2,\dots,2t$, $y_i=j_i\equiv (-j'_i-j''_i)(mod$ $d)$ and
$(j_{2t+1},j_{2t+2},\dots,j_k) \in \Theta \cap \Omega$, we have
\begin{equation}\begin{array}{l}
\lambda_{\overrightarrow{y}}
(\prod\limits_{i=1}^{2t}{\omega^{-x_i(j'_i+j''_i)}})
F(\overrightarrow{j},\overrightarrow{j}',\overrightarrow{j}'')
\otimes
N(\overrightarrow{j}',\overrightarrow{j}'',\overrightarrow{y})
\otimes
C(\overrightarrow{j})\\
=\lambda_{\overrightarrow{y}}
\prod\limits_{i=1}^{2t}{\omega^{x_iy_i}}
h^{j'_1}_1h^{j'_2}_2\dots h^{j'_{2t}}_{2t}
h^{j_1}_1h^{j_2}_2\dots h^{j_{k}}_{k}
h^{j''_1}_1h^{j''_2}_2\dots h^{j''_{2t}}_{2t}\\
=\lambda_{\overrightarrow{y}}
\prod\limits_{i=1}^{2t}{\omega^{x_iy_i}}
h^{j'_1+j_1+j''_1}_1h^{j'_2+j_2+j''_2}_2
\dots h^{j'_{2t}+j_{2t}+j''_{2t}}_{2t}
h^{j_{2t+1}}_{2t+1}\\
h^{j_{2t+2}}_{2t+2}\dots h^{j_{k}}_{k}\\
=\lambda_{\overrightarrow{y}}
\prod\limits_{i=1}^{2t}{\omega^{x_iy_i}}I.
\label{equ:little}
\end{array}\end{equation}
where the first equality makes use of
Eqs.(\ref{equ:h}), (\ref{equ:abc}) and (\ref{equ:efgh}), the
second equality comes from the fact that
$h_1,h_2,\dots,h_{k}$ mutually commute.
So the state of the subsystem $T'_2$ is, up to a normalizing factor,
\begin{equation}\begin{array}{ll}
\widetilde{\rho}^{(T'_2)}&=tr_{T_1,T''_2,M}(\widetilde{\rho})\\
&=\beta \sum\limits_{\overrightarrow{y} \in \mathbb{Z}^{2t}_{d}}
[\lambda_{\overrightarrow{y}}
(\prod\limits_{i=1}^{2t}{\omega^{x_iy_i}})B(\overrightarrow{y})]\\
&=\beta \sum\limits_{\overrightarrow{y} \in \mathbb{Z}^{2t}_{d}}
[\lambda_{\overrightarrow{y}}
(\prod\limits_{i=1}^{2t}{\omega^{x_iy_i}})\bigotimes\limits_{i=1}^t(Z^{y_{2i-1}}X^{y_{2i}})],\\
\label{equ:rhot2}
\end{array}\end{equation}
where $\beta$ is some constant independent of
$\lambda_{\overrightarrow{y}}$.

Finally, after step (3), the state of $T'_2$ becomes, up to a
normalizing factor,
\begin{equation}\begin{array}{l}
V(\overrightarrow{x})\widetilde{\rho}^{(T'_2)}V(\overrightarrow{x})^{\dagger}\\
=\beta \sum\limits_{\overrightarrow{y} \in \mathbb{Z}^{2t}_{d}}
\lambda_{\overrightarrow{y}}
(\prod\limits_{i=1}^{2t}{\omega^{x_iy_i}})
\bigotimes\limits_{i=1}^{t}(Z^{-x_{2i}}X^{x_{2i-1}}Z^{y_{2i-1}}X^{y_{2i}}\\
X^{-x_{2i-1}}Z^{x_{2i}})\\
=\beta \sum\limits_{\overrightarrow{y} \in \mathbb{Z}^{2t}_{d}}
\lambda_{\overrightarrow{y}}
\bigotimes\limits_{i=1}^{t}(Z^{y_{2i-1}}X^{y_{2i}})\\
=\beta \sigma,
\end{array}\end{equation}
where the first equality comes from Eqs.(\ref{equ:vx}) and
(\ref{equ:rhot2}), and the second equality comes from
Eq.(\ref{equ:sigma}). So after this protocol, the final state of
$T'_2$ is exactly the unknown $t$-qudit state $\sigma$.

\hfill $\blacksquare$

\textit{Remark.} It is worth noting that in the above proof the
technique used for proving the validity the teleportation protocol
is different from those used in most literatures. In most previous
work, in order to prove that certain protocols really faithfully
teleport an unknown state, authors usually first restricted the
unknown state to be a pure state, then wrote both the previously
shared entangled state and the unknown state in the vector form,
and finally computed the effect of the protocol on the state
vectors. The calculations were usually very complicated. In
contrast, our approach here is to write the density matrices of
the previously shared entangled state and the unknown state as
linear combinations of generalized Pauli group elements and then take advantage
of their attributes, especially their strong symmetry, to simplify
the calculation. It is entirely possible that this technique
could be applied to a wider class of states besides stabilizer states.

Although theorem 1 only deals with bipartitions, it becomes
the foundation of the following theorem which can deal with
general partition plans.

\begin{theorem}
Suppose $\{T_1,T_2,\dots,T_{m+1}\}$ is a partition of $[1,n]$.
If there exist subgroups $P_1,P_2,\dots,P_{m+1}$ of $S$ such that
\begin{equation}\begin{array}{l}
S^{(T_{m+1})}=\prod\limits_{i=1}^{m+1} P^{(T_{m+1})}_i;\\
P^{(T_{m+1})}_i \cong G^{(d)}_{a_i}, \forall 1 \le i \le m;\\
P^{(T^C_{m+1}-T_i)}_i=\{\gamma^c I\}_{c \in \mathbb{Z}_{2d}}, \forall 1 \le i \le m;\\
P^{(T_{m+1})}_{m+1}\cong\langle \gamma,Z^{c_{1}}_{1},Z^{c_{2}}_{2},
\dots,Z^{c_{s}}_s, X^{d_{1}}_{1},X^{d_2}_{2},\dots,X^{d_{u}}_{u}\rangle,\\
\label{equ:ptcm}
\end{array}\end{equation}
for some $a_1,a_2,\dots,a_m \ge 0$, $s \ge u \ge 0$, and
$c_1,c_2,\dots,c_s, d_1,d_2,\dots,d_u \in \mathbb{Z}_d$,
then $(a_1,a_2,\dots,a_m)$ is an achievable teleportation capacity
for $\rho_S$ with respect to the partition
$\{T_1,T_2,\dots,T_{m+1}\}$.
\end{theorem}

\textit{Proof:}
Define
\begin{equation}
P=\prod\limits_{i=1}^{m}{P_i}.
\end{equation}
Then by Eq.(\ref{equ:ptcm}) we obtain
\begin{equation}
P^{(T_{m+1})} \cong G^{(d)}_b,
\end{equation}
where $b=\sum_{i=1}^m{a_i}$.
So $P$ and $P_{m+1}$ satisfy the condition of theorem 1
with respect to the bipartition $\{T^C_{m+1},T_{m+1}\}$.
Consequently, if the subsystem
$T^C_{m+1}=\bigcup_{i=1}^{m}T_i$ belongs to a single party Alice,
she can faithfully teleport $b$ unknown qudits to Bob who holds
the subsystem $T_{m+1}$. And they can achieve
this by performing the protocol presented in the proof of theorem
1. Actually, we are going to prove that under the given condition
Eq.(\ref{equ:ptcm}), Alice's projective measurement in step (2) in that protocol
can be realized by LOCC with respect to the partition
$\{T_1, T_2, \dots, T_{m}\}$ (at the same time $a_1, a_2, \dots, a_m$ of
the $b$ unknown qudits are also distributed along with
$T_1, T_2, \dots, T_m$ respectively).

Suppose $|T_i|=q_i$, $\forall i=1,2,\dots,m+1$. Also,
define $b_1=0$, $b_i=\sum_{j=1}^{i-1}{a_j}$, $\forall i=2,3,\dots,m+1$.

By Eq.(\ref{equ:ptcm}) we can find independent generators
$g_1,g_2,\dots,g_{k}$ of $S$ such that
$\forall i=1,2,\dots,m$, $\forall j=1,2,\dots,a_i$,
\begin{equation}\begin{array}{l}
g_{2b_i+2j-1}=I^{(T_1)} \otimes \dots \otimes I^{(T_{i-1})} \otimes R_{2b_i+2j-1}
\otimes I^{(T_{i+1})}\\
\otimes \dots \otimes I^{(T_{m})} \otimes \overline{Z}_{b_i+j},\\
g_{2b_i+2j}=I^{(T_1)} \otimes \dots \otimes I^{(T_{i-1})} \otimes R_{2b_i+2j}
\otimes I^{(T_{i+1})}\\
\otimes \dots \otimes I^{(T_{m})} \otimes \overline{X}_{b_i+j};\\
\end{array}\end{equation}
$\forall i=1,2,\dots,s$, $\forall j=1,2,\dots,u$,
\begin{equation}\begin{array}{l}
g_{2b+i}=W_{2b+i} \otimes \overline{Z}^{c_i}_{b+i},\\
g_{2b+s+j}=W_{2b+s+j} \otimes \overline{X}^{d_j}_{b+j};\\
\end{array}\end{equation}
$\forall i=2b+s+u+1,2b+s+u+2,\dots,k$,
\begin{equation}\begin{array}{l}
g_{i}=W_{i} \otimes I^{(T_{m+1})},\\
\end{array}\end{equation}
where $I^{(T_i)}$ is the identity operator on the subsystem $T_i$,
$\forall i=1,2,\dots,m+1$;
$R_{2b_i+1},R_{2b_i+2},\dots,R_{2b_{(i+1)}}$ are some operators
on the subsystem $T_i$, $\forall i=1,2,\dots,m$;
$\overline{Z}_{1},\overline{Z}_{2},\dots,\overline{Z}_{b+s},
\overline{X}_{1},\overline{X}_{2},\dots,\overline{X}_{b+u} \in G^{(d)}_{q_{m+1}}$
are operators on the subsystem $T_{m+1}$;
$W_{2b+1}, W_{2b+2},\dots,W_{k}$ are some operators
on the subsystem $T^C_{m+1}$.

By lemma 1, we can find a unitary operator $U$ acting on $T_{m+1}$ such that
\begin{equation}\begin{array}{ll}
U\overline{Z}_{i}U^{\dagger}=Z_{i},&\forall 1 \le i \le b+s;\\
U\overline{X}_{j}U^{\dagger}=X_{j},&\forall 1 \le j \le b+u.\\
\label{equ:ugu}
\end{array}\end{equation}

Define
\begin{equation}\begin{array}{l}
h_{i}=(I \otimes U)g_{i}(I \otimes U^{\dagger}),\\
\end{array}\end{equation}
$\forall i=1,2,\dots,2b$.
Then we have
$\forall i =1,2,\dots,m$, $\forall j=1,2,\dots,a_i$,
\begin{equation}\begin{array}{l}
h_{2b_i+2j-1}=I^{(T_1)} \otimes \dots \otimes I^{(T_{i-1})} \otimes R_{2b_i+2j-1}
\otimes I^{(T_{i+1})}\\
\otimes \dots \otimes I^{(T_{m})} \otimes {Z}_{b_i+j},\\
h_{2b_i+2j}=I^{(T_1)} \otimes \dots \otimes I^{(T_{i-1})} \otimes R_{2b_i+2j}
\otimes I^{(T_{i+1})}\\
\otimes \dots \otimes I^{(T_{m})} \otimes {X}_{b_i+j}.\\
\label{equ:hij}
\end{array}\end{equation}
Since $g_1,g_2,\dots,g_{2b}$ are commuting operators,
by the definition of $h_1,h_2,\dots,h_{2b}$, we know they are also commuting operators.

Suppose
\begin{equation}
T_{m+1}=\{i_1,i_2,\dots,i_{q_{m+1}}\}
\end{equation}
with $i_1<i_2<\dots<i_{q_{m+1}}$. One can see
Eq.(\ref{equ:hij}) implies $b \le q_{m+1}$.
So for $i=1,2,\dots,m$, define
\begin{equation}\begin{array}{l}
T'_i=\{i_{b_i+1},i_{b_i+2},\dots,i_{b_{(i+1)}}\},\\
Q_i=T_i \cup T'_i.\\
\end{array}\end{equation}
Then let
\begin{equation}\begin{array}{l}
T''=\bigcup_{i=1}^{m}{T'_i}=\{i_1,i_2,\dots,i_b\},\\
T=\bigcup_{i=1}^{m}Q_i=(\bigcup_{i=1}^{m}{T_i}) \cup T''.
\end{array}\end{equation}

Now for $i=1,2,\dots,m$, $j=1,2,\dots,a_i$, define
\begin{equation}\begin{array}{ll}
h'_{2b_i+2j-1}&=I^{(T_1)} \otimes \dots \otimes I^{(T_{i-1})} \otimes R_{2b_i+2j-1}
\otimes I^{(T_{i+1})}\\
&\otimes \dots \otimes I^{(T_{m})} \otimes h^{(T'')}_{2b_i+2j-1}\\
&=I^{(T_1)} \otimes \dots \otimes I^{(T_{i-1})} \otimes R_{2b_i+2j-1}
\otimes I^{(T_{i+1})}\\
&\otimes \dots \otimes I^{(T_{m})} \otimes Z_{b_i+j},\\
h'_{2b_i+2j}&=I^{(T_1)} \otimes \dots \otimes I^{(T_{i-1})} \otimes R_{2b_i+2j}
\otimes I^{(T_{i+1})}\\
&\otimes \dots \otimes I^{(T_{m})} \otimes h^{(T'')}_{2b_i+2j}\\
&=I^{(T_1)} \otimes \dots \otimes I^{(T_{i-1})} \otimes R_{2b_i+2j}
\otimes I^{(T_{i+1})}\\
&\otimes \dots \otimes I^{(T_{m})}  \otimes X_{b_i+j},\\
\label{equ:h2'}
\end{array}\end{equation}
Then $h'_1,h'_2,\dots,h'_{2b}$ are commuting operators on the subsystem $T$.
Moreover, $\forall i=1,2,\dots,m$, $h'_{2b_i+1}, h'_{2b_i+2},\dots,h'_{2b_{(i+1)}}$
only act nontrivially on the subsystem $Q_i$,
i.e.
\begin{equation}
h'^{(Q^C_i)}_{2b_i+1}=h'^{(Q^C_i)}_{2b_i+2}=\dots=h'^{(Q^C_i)}_{2b_{(i+1)}}=I.
\end{equation}

Now for $i =1,2,\dots,m$, $j=1,2,\dots,a_i$, define
\begin{equation}\begin{array}{ll}
h''_{2b_i+2j-1}&=R_{2b_i+2j-1} \otimes h^{(T'_i)}_{2b_i+j-1}\\
&=R_{2b_i+2j-1} \otimes Z_{j},\\
h''_{2b_i+2j}&=R_{2b_i+2j} \otimes h^{(T'_i)}_{2b_i+j}\\
&=R_{2b_i+2j}\otimes X_{j}.\\
\label{equ:h2''}
\end{array}\end{equation}
Then $h''_{2b_i+1},h''_{2b_i+2},\dots,h''_{2b_{(i+1)}}$ are commuting operators
on the subsystem $Q_i$. Furthermore,
$\forall \overrightarrow{x}=(x_1,x_2,\dots,x_{2b}) \in \mathbb{Z}^{2b}_d$,
$\forall \overrightarrow{j}=(j_1,j_2, \dots, j_{2b}) \in \mathbb{Z}^{2b}_d$,
\begin{equation}\begin{array}{l}
\prod\limits_{l=1}^{2b}({{{\omega^{-j_lx_l}h'^{j_l}_l}}})
=\bigotimes\limits_{i=1}^{m}{[\prod\limits_{l=2b_i+1}^{2b_{(i+1)}}{(\omega^{-x_{l}}h''_{l})^{j_{l}}}]}.
\label{equ:sep}
\end{array}\end{equation}
Consequently,
\begin{equation}\begin{array}{l}
P(h'_1,h'_2,\dots,h'_{2b};\overrightarrow{x})\\
=\frac{1}{d^{2b}}\prod\limits_{l=1}^{2b}({{\sum\limits_{j=0}^{d-1}{\omega^{-jx_l}h'^{j}_l}}})\\
=\frac{1}{d^{2b}}\sum\limits_{j_1,\dots,j_{2b}=0}^{d-1}\prod\limits_{l=1}^{2b}({{{\omega^{-j_lx_l}h'^{j_l}_l}}})\\
=\frac{1}{d^{2b}}\sum\limits_{j_1,\dots,j_{2b}=0}^{d-1}
\bigotimes\limits_{i=1}^{m}{[\prod\limits_{l=2b_i+1}^{2b_{(i+1)}}{(\omega^{-x_l}h''_l)^{j_l}}]}\\
=\frac{1}{d^{2b}}\bigotimes\limits_{i=1}^{m}
[\sum\limits_{j_{2b_i+1},\dots,j_{2b_{(i+1)}}=0}^{d-1}{\prod\limits_{l=2b_i+1}^{2b_{(i+1)}}
{(\omega^{-x_l}h''_l)^{j_l}}}]\\
=\bigotimes\limits_{i=1}^{m}[\frac{1}{d^{2a_i}}
{\prod\limits_{l=2b_i+1}^{2b_{(i+1)}}
\sum\limits_{j=0}^{d-1}{(\omega^{-x_l}h''_l)^{j}}}]\\
=\bigotimes\limits_{i=1}^{m}
P(h''_{2b_i+1},h''_{2b_i+2},\dots,h''_{2b_{(i+1)}};\overrightarrow{x}|_{[2b_i+1,2b_{(i+1)}]}),
\end{array}\end{equation}
where $\overrightarrow{x}|_{[2b_i+1,2b_{(i+1)}]}=(x_{2b_i+1},
x_{2b_i+2}, \dots, x_{2b_{(i+1)}})$. The first equality
comes from Eq.(\ref{equ:projection}),
the third equality comes from Eq.(\ref{equ:sep}),
the fifth equality makes use of $b=\sum_{i=1}^m{a_i}$,
and the last equality also comes from Eq.(\ref{equ:projection}). So
$P(h'_1,h'_2,\dots,h'_{2b};\overrightarrow{x})$ is simply the
tensor product of the projection operators
$P(h''_{2b_i+1},h''_{2b_i+2},\dots,h''_{2b_{(i+1)}};\overrightarrow{x}|_{[2b_i+1,2b_{(i+1)}]})$
on each subsystem $Q_i$.

Therefore, by making a little modification to the protocol in the
proof of theorem 1, we get the protocol for our teleportation with
respect to the partition $\{T_1,T_2, \dots, T_{m+1}\}$ as follows:

(1)$A_{m+1}$ performs the unitary operation $U$ on the subsystem
$T_{m+1}$.

(2)Suppose $A_i$ has an unknown $a_i$-qudit state $\sigma_i$,
$\forall i =1,2,\dots,m$. $A_i$ performs the projective measurement
consisting of the projection operators
$\{P(h''_{2b_i+1},h''_{2b_i+2},\dots,h''_{2b_{(i+1)}};\overrightarrow{x})\}_
{\overrightarrow{x} \in \mathbb{Z}^{2a_i}_d}$ on his $T_i$ subsystem of $\rho_S$ and
$\sigma_i$, and then tells the measurement outcome
$\overrightarrow{x}=(x_{2b_i+1},
x_{2b_i+2},\dots,x_{2b_{(i+1)}})$ to $A_{m+1}$, $\forall i =1,2,\dots,m$.

(3)$A_{m+1}$ performs the unitary operation
\begin{equation}
V(\overrightarrow{x})=
\bigotimes\limits_{i=1}^{b}(Z^{-x_{2i}}X^{x_{2i-1}})
\end{equation}
on the subsystem $\bigcup_{i=1}^{m}T'_i$.

Then by the proof of the theorem 1, we know that after this
protocol, the final states of $T'_1, T'_2, \dots, T'_m$ become
$\sigma_1, \sigma_2, \dots, \sigma_m$ respectively.

\hfill $\blacksquare$

\textit{Remark 1.} One can easily see that in the two protocols
presented in the proofs of theorem 1 and 2, the receiver can
actually perform the unitary operation $U$ after receiving the
senders' measurement outcomes, i.e. the order of step (1) and (2)
can be altered.

\textit{Remark 2.} One can see that our two theorems above do not
require the state $\rho_S$ to be a pure stabilizer state. When $S$
is an incomplete stabilizer, $\rho_S$ is a mixed state. In the
subsequent section, we will also give concrete examples of mixed
stabilizer states which are useful for perfect teleportation, even
with respect to several different partition plans. Our argument
mainly depends on the structure of the restrictions of
$S$ on each subsystem $T_i$. The purity of
$\rho_S$ is not an essential property that can greatly influence
its teleportation capability.

\section{Illustrations}

In this section we will analyze several states by using our
theorems. In each example, the matrices $X$ and $Z$ are $X_{(d)}$
and $Z_{(d)}$ defined by Eq.(\ref{equ:pauli}) with the
corresponding dimension $d$. We also use the notation $X_j$
denotes the operation $X$ acting on the $j$th qudit and similarly
for $Z_j$.

We will consider three examples. The first example is
re-examination of the standard teleportation protocol from our
perspective. The second and third examples are detailed
illustrations of how to find the achievable teleportation capacity
and construct the corresponding protocol by utilizing our two
theorems. The third example also proves the existence of mixed
stabilizer states which are useful for perfect teleportation.

\begin{example}
Let us begin with the standard teleportation protocol. Let
\begin{equation}
|\Phi^{+}\rangle=\frac{1}{\sqrt{d}}\sum\limits_{i=0}^{d-1}{|ii\rangle}
\end{equation}
be the maximally entangled state in the $d \times d$ system. It is
a stabilizer state and its stabilizer is $S=\langle
g_1,g_2\rangle$, where
\begin{equation}\begin{array}{l}
g_1=Z^{-1}_1Z_2,\\
g_2=X_1X_2.
\end{array}\end{equation}
Consider the partition $\{\{1\},\{2\}\}$. We have $g_1^{(\{2\})}=Z$,
$g_2^{(\{2\})}=X$ and consequently $S^{(\{2\})}\cong G^{(d)}_1$.
So by theorem 1, if Alice and Bob hold the first and second qudits of
$|\Phi^{+}\rangle$ respectively, then Alice can faithfully
teleport an unknown qudit state to Bob. Moreover, in this special
case, the protocol presented in the proof of theorem 1 becomes:
Alice first performs the projective measurement in the basis of
the simultaneous eigenstates of $g_1, g_2$ on her subsystem of
$|\Phi^{+}\rangle$ and the unknown qudit; if her measurement
outcome corresponds to the eigenvalues $\omega^{a},\omega^{b}$ of
$g_1, g_2$ for some $a,b \in \mathbb{Z}_d$, then Bob performs the
unitary operation $Z^{-b}X^{a}$ on his qudit. One can easily see that this
protocol is exactly the standard teleportation protocol.
\end{example}

\begin{example}
Consider a five-qutrit system, i.e. $d=3$, $n=5$. Define
\begin{equation}\begin{array}{l}
g_1=X_1X^2_2X_3Z_4Z_5,\\
g_2=Z^2_1Z_2I_3X_4I_5,\\
g_3=Z_1Z_2Z_3I_4X_5,\\
g_4=X_1X_2Z_3X_4Z^2_5,\\
g_5=I_1I_2Z^2_3X^2_4I_5.
\end{array}\end{equation}
They are five independent commuting operators in $G'^{(3)}_5$.
Then
\begin{equation}\begin{array}{l}
S=\langle g_1,g_2,g_3,g_4,g_5\rangle
\end{array}\end{equation}
is a complete stabilizer.
Suppose $|\psi_S\rangle$ is the pure state stabilized by $S$.
Then
\begin{equation}
\rho_S=|\psi_S\rangle\langle\psi_S|=\frac{1}{3^5}\prod\limits_{i=1}^{5}(\sum\limits_{j=0}^{2}{g_i^j}).
\end{equation}

Consider the partition $\{T_1=\{1,2\},T_2=\{3,4,5\}\}$. We have
\begin{equation}\begin{array}{l}
g^{(T_2)}_1=X \otimes Z \otimes Z,\\
g^{(T_2)}_2=I \otimes X \otimes I,\\
g^{(T_2)}_3=Z \otimes I \otimes X,\\
g^{(T_2)}_4=Z \otimes X \otimes Z^2,\\
g^{(T_2)}_5=Z^2 \otimes X^2 \otimes I.\\
\end{array}\end{equation}
One can check that we can write $g^{(T_2)}_1=\overline{Z}_1$,
$g^{(T_2)}_2=\overline{X}_1$, $g^{(T_2)}_3=\overline{Z}_2$,
$g^{(T_2)}_4=\overline{X}_2$,
$g^{(T_2)}_5=\overline{Z}_3$.
Let
\begin{equation}\begin{array}{l}
P_1=\langle g_1,g_2,g_3,g_4\rangle,\\
P_2=\langle g_5\rangle.
\end{array}\end{equation}
Then
\begin{equation}\begin{array}{l}
S^{(T_2)}=P^{(T_2)}_1P^{(T_2)}_2,\\
P^{(T_2)}_1 \cong G^{(3)}_2,\\
P^{(T_2)}_2 \cong \langle \gamma,Z \rangle,\\
\end{array}\end{equation}
where $\gamma=e^{i\frac{\pi}{3}}$. So by theorem 1, if Alice and
Bob hold the $T_1$ and $T_2$ subsystems of $\rho_S$ respectively,
then Alice can faithfully teleport an unknown two-qutrit state
$\sigma$ to Bob. We now show how to construct the corresponding
teleportation protocol. By lemma 1 and its proof, we can find a
unitary operation $U$ acting on $T_2$ such that
$Ug^{(T_2)}_1U^{\dagger}=Z_1$, $Ug^{(T_2)}_2U^{\dagger}=X_1$,
$Ug^{(T_2)}_3U^{\dagger}=Z_2$, $Ug^{(T_2)}_4U^{\dagger}=X_2$,
$Ug^{(T_2)}_5U^{\dagger}=Z_3$. Define
\begin{equation}\begin{array}{l}
h_1=(I \otimes U)g_1(I \otimes U)^{\dagger}=X \otimes X^2 \otimes Z \otimes I \otimes I,\\
h_2=(I \otimes U)g_2(I \otimes U)^{\dagger}=Z^2 \otimes Z \otimes X \otimes I \otimes I,\\
h_3=(I \otimes U)g_3(I \otimes U)^{\dagger}=Z \otimes Z \otimes I \otimes Z \otimes I,\\
h_4=(I \otimes U)g_4(I \otimes U)^{\dagger}=X \otimes X \otimes I \otimes X \otimes I.\\
\end{array}\end{equation}
Let $T'_2=\{3,4\}$. Then define
\begin{equation}\begin{array}{ll}
h'_1&=X \otimes X^2 \otimes h^{(T'_2)}_1\\
&=X \otimes X^2 \otimes Z \otimes I,\\
h'_2&=Z^2 \otimes Z \otimes h^{(T'_2)}_2\\
&=Z^2 \otimes Z \otimes X \otimes I,\\
h'_3&=Z \otimes Z \otimes h^{(T'_2)}_3\\
&=Z \otimes Z \otimes I \otimes Z,\\
h'_4&=X \otimes X \otimes h^{(T'_2)}_4\\
&=X \otimes X \otimes I \otimes X.\\
\end{array}\end{equation}
The protocol is as follows: (1)Bob performs the unitary operation
$U$ on the $T_2$ subsystem of $\rho_S$. (2)Alice performs the
projective measurement in the basis of the simultaneous
eigenstates of $h'_1,h'_2,h'_3,h'_4$ on her $T_1$ subsystem of
$\rho_S$ and $\sigma$. Suppose the measurement outcome corresponds
to the eigenvalues
$\omega^{x_1},\omega^{x_2},\omega^{x_3},\omega^{x_4}$ of
$h'_1,h'_2,h'_3,h'_4$, where $\omega=e^{i\frac{2\pi}{3}}$. She
tells $x_1,x_2,x_3,x_4$ to Bob. (3)Bob performs the unitary
operation $V=Z^{-x_2}X^{x_1} \otimes Z^{-x_4}X^{x_3}$ on the
$T'_2$ subsystem. Then after this procedure, Bob's $T'_2$
subsystem becomes the state $\sigma$.
\end{example}

\begin{example}
Consider an eight-qubit system. Define
\begin{equation}\begin{array}{l}
g_1=X_1Y_2I_3I_4I_5Z_6Y_7I_8,\\
g_2=X_1Z_2I_3I_4I_5X_6Y_7I_8,\\
g_3=I_1I_2Z_3Y_4Z_5I_6Y_7X_8,\\
g_4=I_1I_2Z_3I_4X_5I_6Y_7Z_8,\\
g_5=I_1I_2Z_3Z_4X_5Y_6X_7Y_8,\\
g_6=I_1I_2X_3X_4Z_5Y_6Z_7Y_8,\\
g_7=Z_1X_2I_3Z_4X_5I_6I_7I_8.\\
\end{array}\end{equation}
They are seven commuting operators in $G'^{(2)}_8$. Let
\begin{equation}\begin{array}{l}
S=\langle g_1,g_2,\dots,g_7\rangle.
\end{array}\end{equation}
Then the maximally mixed
state over the subspace stabilized by $S$ is
\begin{equation}
\rho_S=\frac{1}{2^8}\prod\limits_{i=1}^7(I+g_i).
\end{equation}

Consider the partition
$\{T_1=\{1,2\},T_2=\{3,4,5\}$,
$T_3=\{6,7,8\}\}$. We have
\begin{equation}\begin{array}{l}
g^{(T_3)}_1=Z \otimes Y \otimes I,\\
g^{(T_3)}_2=X \otimes Y \otimes I,\\
g^{(T_3)}_3=I \otimes Y \otimes X,\\
g^{(T_3)}_4=I \otimes Y \otimes Z,\\
g^{(T_3)}_5=Y \otimes X \otimes Y,\\
g^{(T_3)}_6=Y \otimes Z \otimes Y.\\
g^{(T_3)}_7=I \otimes I \otimes I,\\
g^{(T_2)}_1=g^{(T_2)}_2=I \otimes I \otimes I,\\
g^{(T_1)}_3=g^{(T_1)}_4=g^{(T_1)}_5=g^{(T_1)}_6=I \otimes I.\\
\end{array}\end{equation}
One can check that we can write $g^{(T_3)}_1=\overline{Z}_1$,
$g^{(T_3)}_2=\overline{X}_1$, $g^{(T_3)}_3=\overline{Z}_2$,
$g^{(T_3)}_4=\overline{X}_2$, $g^{(T_3)}_5=\overline{Z}_3$,
$g^{(T_3)}_6=\overline{X}_3$.
Let
\begin{equation}\begin{array}{l}
P_1=\langle g_1,g_2\rangle,\\
P_2=\langle g_3,g_4,g_5,g_6\rangle,\\
P_3=\langle g_7\rangle.\\
\end{array}\end{equation}
Then
\begin{equation}\begin{array}{l}
S^{(T_3)}=\prod\limits_{i=1}^3P^{(T_3)}_i,\\
P^{(T_3)}_1 \cong G^{(2)}_1,\\
P^{(T_3)}_2 \cong G^{(2)}_2,\\
P^{(T_2)}_1=\{i^cI \otimes I \otimes I\}_{c \in \mathbb{Z}_4},\\
P^{(T_1)}_2=\{i^cI \otimes I\}_{c \in \mathbb{Z}_4},\\
P^{(T_3)}_3=\{i^cI \otimes I \otimes I\}_{c \in \mathbb{Z}_4}.\\
\end{array}\end{equation}
So by theorem 2, $(1,2)$ is an achievable teleportation capacity for
$\rho_S$ with respect to the partition
$\{\{1,2\},\{3,4,5\},\{6,7,8\}\}$. In other words, supposing
Alice, Bob and Charlie hold the subsystems $\{1,2\}$, $\{3,4,5\}$
and $\{6,7,8\}$ of $\rho_S$ respectively, if Alice has an unknown
qubit state $\sigma_1$ and Bob has an unknown two-qubit state
$\sigma_2$, then they can simultaneously faithfully teleport
$\sigma_1$ and $\sigma_2$ to Charlie. We now show how to construct
the corresponding teleportation protocol. By lemma 1 and its
proof, we can find a unitary operation $U$ acting on $T_3$ such
that $Ug^{(T_3)}_1U^{\dagger}=Z_1$, $Ug^{(T_3)}_2U^{\dagger}=X_1$,
$Ug^{(T_3)}_3U^{\dagger}=Z_2$, $Ug^{(T_3)}_4U^{\dagger}=X_2$,
$Ug^{(T_3)}_5U^{\dagger}=Z_3$, $Ug^{(T_3)}_6U^{\dagger}=X_3$.
Define
\begin{equation}\begin{array}{l}
h_1=(I \otimes U)g_1(I \otimes U)^{\dagger}=X_1Y_2I_3I_4I_5Z_6I_7I_8,\\
h_2=(I \otimes U)g_2(I \otimes U)^{\dagger}=X_1Z_2I_3I_4I_5X_6I_7I_8,\\
h_3=(I \otimes U)g_3(I \otimes U)^{\dagger}=I_1I_2Z_3Y_4Z_5I_6Z_7I_8,\\
h_4=(I \otimes U)g_4(I \otimes U)^{\dagger}=I_1I_2Z_3I_4X_5I_6X_7I_8,\\
h_5=(I \otimes U)g_5(I \otimes U)^{\dagger}=I_1I_2Z_3Z_4X_5I_6I_7Z_8,\\
h_6=(I \otimes U)g_6(I \otimes U)^{\dagger}=I_1I_2X_3X_4Z_5I_6I_7X_8.
\end{array}\end{equation}
Let $T'_1=\{6\},T'_2=\{7,8\}$. Then define
\begin{equation}\begin{array}{ll}
h''_1&=X \otimes Y \otimes h^{(T'_1)}_1\\
&=X \otimes Y \otimes Z,\\
h''_2&=X \otimes Z \otimes h^{(T'_1)}_2\\
&=X \otimes Z \otimes X,\\
h''_3&=Z \otimes Y \otimes Z \otimes h^{(T'_2)}_3\\
&=Z \otimes Y \otimes Z \otimes Z \otimes I,\\
h''_4&=Z \otimes I \otimes X \otimes h^{(T'_2)}_4\\
&=Z \otimes I \otimes X \otimes X \otimes I,\\
h''_5&=Z \otimes Z \otimes X \otimes h^{(T'_2)}_5\\
&=Z \otimes Z \otimes X \otimes I \otimes Z,\\
h''_6&=X \otimes X \otimes Z \otimes h^{(T'_2)}_6\\
&=X \otimes X \otimes Z \otimes I \otimes X.\\
\end{array}\end{equation}
The protocol is as follows: (1)Charlie performs the unitary
operation $U$ on the subsystem $T_3$ of $\rho_S$. (2.1)Alice
performs the projective measurement consisting of the projection
operators $\{P(h''_1,h''_2;\overrightarrow{x}):
\overrightarrow{x} \in \mathbb{Z}^2_2\}$ on her $T_1$ subsystem of
$\rho_S$ and $\sigma_1$, and then tells the measurement outcome
$\overrightarrow{x}=(x_1,x_2)$ to Charlie; (2.2)Bob performs the
projective measurement consisting of the projection operators
$\{P(h''_3,h''_4, h''_5,h''_6;\overrightarrow{x}): \overrightarrow{x} \in \mathbb{Z}^4_2\}$ on
his $T_2$ subsystem of $\rho_S$ and $\sigma_2$, and then tells the
measurement outcome $\overrightarrow{x}=(x_3,x_4,x_5,x_6)$ to
Charlie; (3)Charlie performs the unitary operation
$V=Z^{-x_2}X^{x_1} \otimes Z^{-x_4}X^{x_3} \otimes
Z^{-x_6}X^{x_5}$ on the subsystem $T'_1 \cup T'_2$. After this
procedure, Charlie's $T'_1$ and $T'_2$ subsystems become the states
$\sigma_1$ and $\sigma_2$ respectively.

Now consider another partition $\{T_1=\{1,6\},T_2=\{3,8\},
T_3=\{2,4,5,7\}\}$. Define
\begin{equation}\begin{array}{l}
g'_5=g_1g_2g_3g_4g_5=-I_1X_2Z_3X_4Z_5I_6X_7I_8,\\
g'_6=g_1g_2g_6=-I_1X_2X_3X_4Z_5I_6Z_7Y_8,\\
g'_7=g_1g_2g_7=-Z_1I_2I_3Z_4X_5Y_6I_7I_8.
\end{array}\end{equation}
Then
\begin{equation}\begin{array}{ll}
S&=\langle g_1,g_2,g_3,g_4,g_5,g_6,g_7\rangle\\
&=\langle g_1,g_2,g_3,g_4,g'_5,g'_6,g'_7 \rangle.
\end{array}\end{equation}
Moreover, we have
\begin{equation}\begin{array}{l}
g^{(T_3)}_1=Y \otimes I \otimes I \otimes Y,\\
g^{(T_3)}_2=Z \otimes I \otimes I \otimes Y,\\
g^{(T_3)}_3=I \otimes Y \otimes Z \otimes Y,\\
g^{(T_3)}_4=I \otimes I \otimes X \otimes Y,\\
g'^{(T_3)}_5=X \otimes X \otimes Z \otimes X,\\
g'^{(T_3)}_6=X \otimes X \otimes Z \otimes Z,\\
g'^{(T_3)}_7=I \otimes Z \otimes X \otimes I.\\
g^{(T_2)}_1=g^{(T_2)}_2=I \otimes I,\\
g^{(T_1)}_3=g^{(T_1)}_4=I \otimes I.\\
\end{array}\end{equation}
One can check that we can write $g^{(T_3)}_1=\overline{Z}_1$,
$g^{(T_3)}_2=\overline{X}_1$, $g^{(T_3)}_3=\overline{Z}_2$,
$g^{(T_3)}_4=\overline{X}_2$,
$g'^{(T_3)}_5=\overline{Z}_3$, $g'^{(T_3)}_6=\overline{X}_3$,
$g'^{(T_3)}_7=\overline{Z}_4$. Let
\begin{equation}\begin{array}{l}
P_1=\langle g_1,g_2\rangle,\\
P_2=\langle g_3,g_4\rangle,\\
P_3=\langle g'_5,g'_6,g'_7\rangle.
\end{array}\end{equation}
Then
\begin{equation}\begin{array}{l}
S^{(T_3)}=\prod\limits_{i=1}^{3}P^{(T_3)}_i,\\
P_1^{(T_3)}\cong P_2^{(T_3)}\cong G^{(2)}_1,\\
P^{(T_2)}_1=P^{(T_1)}_2=\{i^cI \otimes I\}_{c \in \mathbb{Z}_4},\\
P_3^{(T_3)} \cong \langle i,Z_1,X_1,Z_2\rangle.
\end{array}\end{equation}
Therefore, by theorem 2, $(1,1)$ is an achievable
teleportation capacity for $\rho_S$ with respect to the partition
$\{\{1,6\},\{3,8\},\{2,4,5,7\}\}$. In other words, supposing
Alice, Bob and Charlie hold the subsystems $\{1,6\}$, $\{3,8\}$
and $\{2,4,5,7\}$ of $\rho_S$ respectively, if Alice has an
unknown qubit state $\sigma_1$ and Bob has an unknown qubit state
$\sigma_2$, then they can simultaneously faithfully teleport
$\sigma_1$ and $\sigma_2$ to Charlie. The reader can build the
corresponding protocol through an analysis similar to the one
above.

Note that $S=\langle g_1,g_2,\dots,g_7\rangle$ is an incomplete
stabilizer. So $\rho_S$ is a mixed stabilizer state. But it is
still useful for perfect teleportation with respect to at least
two different partition plans.
\end{example}

\section{Conclusion}
In sum, we have studied the possibility of performing perfect
many-to-one teleportation with a previously shared stabilizer
state. We present two sufficient conditions for a stabilizer state
to achieve a given nonzero teleportation capacity with respect to
a given partition plan. The corresponding protocols are also
explicitly constructed. Mixed stabilizer states are also found to
be useful for perfect many-to-one teleportation. Our work provides
a new perspective from the stabilizer formalism to view the
standard teleportation protocol and also suggests a new technique
to analyze the teleportation capability of multipartite entangled
states.

We would like to point out several directions for future
investigations. Firstly, we do not know whether the conditions of
the two theorems are also necessary for $\rho_S$ to achieve a
nonzero capacity with respect to a given partition. We believe
that it is the structure of the restrictions of the stabilizer $S$
on each subsystem that determines teleportation capability of
$\rho_S$. But it seems not easy to reach a thorough understanding.
Secondly, we expect our techniques in the proof of theorem 1 to be
extended to a wider class of entangled states besides stabilizer
states. We think that as long as the considered state exhibits
strong symmetry, our techniques can be readily applied. We hope
our work can stimulate further research on the usefulness of a
general multipartite entangled state for faithful teleportation.

\section*{Acknowledgement}
This work was partly supported by the Natural Science Foundation
of China (Grant Nos. 60621062 and 60503001) and the Hi-Tech
Research and Development Program of China (863 project) (Grant No.
2006AA01Z102).


\begin{thebibliography}{99}

\bibitem{BB93} C. H. Bennett, G. Brassard, C. Crepeau, R. Jozsa, A. Peres, and W. K. Wootters, Phys. Rev. Lett. \textbf{70}, 1895 (1993).

\bibitem{HH99} M. Horodecki, P. Horodecki, R. Horodecki, Phys. Rev. A \textbf{60}, 1888 (1999).

\bibitem{LK00} J. Lee and M. S. Kim, Phys. Rev. Lett. \textbf{84}, 4236 (2000).

\bibitem{B00} K. Banaszek, Phys. Rev. A, \textbf{62}, 024301 (2000).

\bibitem{RH01} J. \v{R}eh\'{a}\v{c}ek, Z. Hradil, J. Fiur\'{a}\v{s}ek, and \v{C}. Brukner, Phys. Rev. A \textbf{64}, 060301(R) (2001).

\bibitem{SL01} W. Son, J. Lee, M. S. Kim and Y. J. Park, Phys. Rev. A \textbf{64}, 064304 (2001).

\bibitem{AF02} S. Albeverio, S. M. Fei and W. L. Yang, Phys. Rev. A \textbf{66}, 012301 (2002).

\bibitem{GK02} S. Ghosh, G. Kar, A. Roy, D. Sarkar and U. Sen,Phys. Rev. A \textbf{66}, 024301 (2002).

\bibitem{VV03} F. Verstraete and H. Verschelde, Phys. Rev. Lett. \textbf{90}, 097901 (2003).

\bibitem{Y03} Ye Yeo, Phys. Rev. A \textbf{67}, 054304 (2003).

\bibitem{RD03} L. Roa, A. Delgado and I. Fuentes-Guridi, Phys. Rev. A \textbf{68}, 022310 (2003).

\bibitem{KC04} H. Kim, Y. W. Cheong and H. W. Lee, Phys. Rev. A \textbf{70}, 012309 (2004).

\bibitem{G04} G. Gour, Phys. Rev. A \textbf{70}, 042301 (2004).

\bibitem{JF05} L. Mi\v{s}ta, Jr. and R. Filip, Phys. Rev. A \textbf{71}, 022319 (2005).

\bibitem{LC05} L. Li and Z. B. Chen, Phys. Rev. A \textbf{72}, 014302 (2005).

\bibitem{GR06} G. Gordon and G. Rigolin, Phys. Rev. A \textbf{73}, 042309 (2006).

\bibitem{BS06} S. Bandyopadhyay and B. C. Sanders, Phys. Rev. A \textbf{74}, 032310 (2006).

\bibitem{M06} L. Masanes, Phys. Rev. Lett. \textbf{96}, 150501 (2006).


\bibitem{LM02} J. Lee, H. Min and S. D. Oh, Phys. Rev. A \textbf{66}, 052318 (2002).

\bibitem{FL03} J. Fang, Y. Lin, S. Zhu and X. Chen, Phys. Rev. A \textbf{67}, 014305 (2003).

\bibitem{LJ05} S. Lee, J. Joo and J. Kim, Phys. Rev. A \textbf{72}, 024302 (2005).

\bibitem{R05} G. Rigolin, Phys. Rev. A \textbf{71}, 032303 (2005).

\bibitem{YC06} Y. Yeo and W. K. Chua, Phys. Rev. Lett. \textbf{96}, 060502 (2006).

\bibitem{CZ06} P. X. Chen, S. Y. Zhu and G. C. Guo, Phys. Rev. A \textbf{74}, 032324 (2006).

\bibitem{AP06} P. Agrawal and A. Pati, Phys. Rev. A \textbf{74}, 062320 (2006).

\bibitem{MX07} Z. X. Man, Y. J. Xia and N. B. An, Phys. Rev. A \textbf{75}, 052306 (2007).

\bibitem{LJ07} S. Lee, J. Joo and J. Kim, Phys. Rev. A \textbf{76}, 012311 (2007).


\bibitem{S95} P.W. Shor, Phys. Rev. A \textbf{52}, R2493 (1995).

\bibitem{S96} A. M. Steane, Phys. Rev. A \textbf{54}, 4741 (1996).

\bibitem{RB01} R. Raussendorf and H. J. Briegel, Phys. Rev. Lett. \textbf{86}, 5188 (2001).

\bibitem{G96} D. Gottesman, Phys. Rev. A \textbf{54}, 1862 (1996).

\bibitem{G97} D. Gottesman, Ph.D. thesis, California Institute of Technology, Pasadena, CA, 1997.

\bibitem{ND07} M. Van den Nest, W. D\"ur and H. J. Briegel, Phys. Rev. Lett. \textbf{98}, 117207 (2007).

\bibitem{TG05} G. T\'{o}th and O. G\"uhne, Phys. Rev. A \textbf{72}, 022340 (2005).

\bibitem{WY07} G. Wang, M. Ying, Phys. Rev. A \textbf{75}, 052332 (2007).

\bibitem{G98} D. Gottesman, in Quantum Computing and Quantum Communications:
First NASA International Conference, edited by C. P. Williams
~Springer-Verlag, Berlin, 1999.

\bibitem{HD05} E. Hostens, J. Dehaene and B. De Moor, Phys. Rev. A \textbf{71}, 042315 (2005).

\bibitem{Y02} A. Y. Vlasov, e-print quant-ph/0210049.

\bibitem{NB02} M. A. Nielsen, M. J. Bremner, J. L. Dodd, A. M. Childs, and
C. M. Dawson, Phys. Rev. A \textbf{66}, 022317 (2002).


\bibitem{BD06} T. Brun, I. Devetak and M. H. Hsieh, Science, \textbf{314}, 436 (2006).

\end{thebibliography}
\end{document}